# Knowledge-Driven Game Design by Non-Programmers

**Iaakov Exman and Avinoam Alfia**

Software Engineering Department
The Jerusalem College of Engineering - Azrieli
POB 3566, Jerusalem, 91035, Israel
`iaakov@jce.ac.il, avinoama2@gmail.com`

**Abstract.** Game extension is an entertaining activity that offers an opportunity to test new design approaches by non-programmers. The real challenge is to enable this activity by means of a suitable infrastructure. We propose a knowledge-driven approach with natural game-player concepts. These concepts, found in game ontologies, include game abstractions and rules for game moves. The approach has been implemented and tested for *board* games. These include tic-tac-toe as a simplest example, enabling extensions of tic-tac-toe, say to a four-by-four board and Sudoku, a single player game of a very different nature.

## 1  Introduction

Software games have virtualized and borrowed ideas from material board games, and have enabled development of its own games and styles. Software game extension is a very entertaining and challenging activity. In this work we propose an infrastructure for extensible software games for non-programmers.

This work has a double motivation:

- *Software Engineering* – starting from domain knowledge to support automatic generation of software by non-programmers;
- *Games' Domain* – characterization of the specific domain of board games in terms of sets of basic concepts, states and transition rules.

To state the infrastructure requirements, one needs to clarify what is meant by non-programmers. Following Exman [2], a non-programmer is not just an ignorant of JAVA and other programming languages. A non-programmer, especially in our knowledge oriented context, does have characteristics in common to programmers, such as the ability to formulate domain knowledge. In particular, one may assume the capability of formulating or at least reading an ontology. Therefore, the infrastructure is knowledge-driven, viz. it is based upon player's knowledge about games – embodied in game ontologies. Basic concepts for game design and extension are game abstractions such as the concept of a game-board and rules for game moves.

This work describes the extensible game infrastructure architecture, its implementation and uses a few case studies as demonstration of the approach.

## 1.1   Related Work

Here we present a concise review of related work. Game tools have been especially developed for non-programmers. For instance, Brom and collaborators [1] discuss agent based systems for non-experts, say students and other non-programmers.

Games have also been regarded as educational tools. Johnson and Beal [5] apply games to language learning. A similar context is games for and by children. Good and Robertson [3] discuss the effects of games on learning and skill development, by means of computer games authored by children.

McNaughton and collaborators [6] take a software engineering oriented approach to look at games. They discuss generative design patterns for role playing games. Moreno-Ger et al. [7] describe a documental approach to computer game development. The games in the referred paper – adventure games with relatively complex and variable user interfaces – differ from the board-games in our own work.

The knowledge-driven line of research for games is represented by various game ontologies as a basis for game design. See e.g. the references by Hagen [4], Studer et al. [8], and the Game Ontology Project by Zagal et al. [9], [10], [11].

In the remaining of the paper we introduce games as knowledge-ware (section 2), describe rules and rule-sets (section 3), overview the software architecture and implementation of our tool (section 4), discuss case studies as a demonstration of the approach (section 5) and conclude with a discussion (section 6).

## 2  Games as Knowledge-Ware

Here we characterize games from a knowledge point of view. We specifically refer to board games, which we have chosen as the main subject matter of this work.

We can divide the characterization into two principal parts: a- the overall game concepts; b- the rules of the game that characterize the players' behavior. An important organized source of concepts is the set of proposed game ontologies.

### 2.1    Abstract Game Characterization

High-level concepts needed for game design infra-structure include:

- *Game* – an abstraction of the Game activity;
- *Game-board* – a usually 2-dimensional mathematical matrix with integer values (that in principle can be recasted to any type);
- *Player* – a human or robot participant in the game;
- *Owning* – a relation between two different sets, for instance a player owns a tile in the board game. Owning does not necessarily imply a specific semantic content, see e.g. [11];

For lower-level concepts, one conceivably has various alternatives. It seems that definitive ontologies in the game context still are an open issue.

For instance, in reference [8] one finds a compact ontology for their board-game method, which is divided in:

- *Global Definitions* – such as movable objects, states;
- *Input* – such as pieces, locations, initial_states, moves, legal state;
- *Ouput* – goal_states;
- *Internal Definitions* – such as current_states, potential_successor_states.

### 2.2    Logical Rules for Game Player Behavior

Rules are planned to work in an Event-Condition-Action cycle. If an event occurs caused by a game player, a condition is evaluated, and if it is satisfied, an action is triggered.

Rules can be classified into generic rules for all games and specific rules for particular games. Generic rules for all games comprise those concerning with:

- *Game Start* – refer to game instance creation, board initial conditions and initial players joining the game;
- *Game Termination* – refer to periodic checking of termination conditions and effective termination declaration;
- *Mid-Game Moves* – general events and actions relevant after the game starts and before it terminates.

# 3 Rules and Rule-Sets

Here we explain examples of generic rules – simple and rule sets.

### 3.1   Simple Rules

Simple rules have a single condition to be evaluated and simple actions, and no components. An example is the generic *Game Start* rule:

- *Event*–Game Start;
- *Condition*– check the type of game (to set the relevant actions);
- *Action* – change game state to *started*, send message to users, tile listen to click events.

### 3.2   Rule Composition into Rule-Sets

To facilitate game understanding, one uses hierarchical composition of rules, like software composition. Components are rules or actions triggered by other rules.

An example is the generic *Tile Click* rule. When a tile is clicked, one checksif the game is finished, i.e. there is a winner, or otherwise switch the player. These are generic components best defined as separate rules and called by *Tile Click*:

- *Event* – Tile Click;
- *Condition* – check the nature of the game; check if the game is not terminated; check if this is the current player turn; check if the tile is not taken;
- *Action* – tile set owner;
- *Component rules* – Rule set: "check winner"; Rule: "Switch player".

Another example is the "check winner" rule composed of elementary tile checks or itself defined as a whole pattern. For instance, tic-tac-toe checks rows, columns and diagonals in the board. Each of these requires checking three aligned tiles.

# 4 Architecture and Implementation of the *GAMES* Tool

A tool called *GAMES* – for Game Management System – was designed to implement the Knowledge-Driven game design and extension approach.

The *GAMES* software architecture is reflected in the system behavior shown in the statechart in Fig. 1.It has two upper modules: server and client. These modules communicate by means of the GameUI (User Interface) to/from the Request-Handler. The client GameBoard is a faithful copy of the server GameBoard.

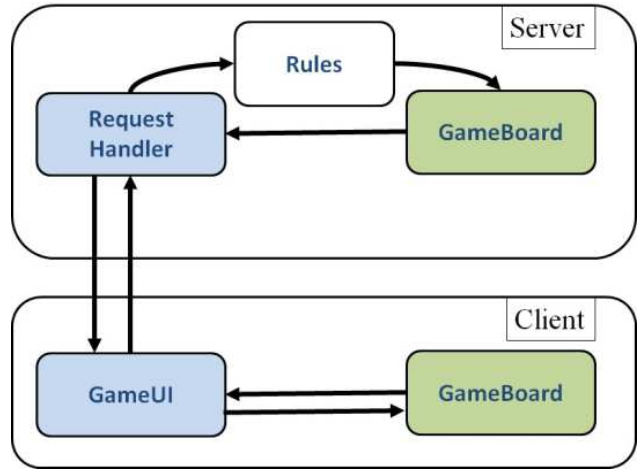

**Fig. 1.** *GAMES* Architecture. This statechart shows states for two modules: server and client.

Interactions in the Server are as follows: an event in the Request Handler causes evaluation of Rules; the latter may trigger an action in the GameBoard that in turn creates a command to be sent back to the clients.

### 4.1   *GAMES* Implementation

*GAMES* was implemented in Drupal – a widespread usage CMS (Content Management System). Besides its core modules, we used a Rules module to implement rules that we created for our games. The Drupal core defines data structures as *entities*.

We created a set of basic entities. The most basic entity is *game*. A *board game* is a set of game attributes. To make the game playable there is a state machine entity called *running game*. The latter describes and saves the current state of each game.

For the sake of efficiency the underlying ontology is implicit as the set of entities and their attributes. In contrast, rules are explicit in the *rules* module, which evaluate event conditions and actions to perform, when the player triggers events.

The player interface is a browser independent web-service client. For illustration, Fig. 2 shows a screen print of the *GAMES* player interface.

The rules editor interface is located in the back-end of the tool. It can be accessed only with the correct permissions. See for instance Fig. 3, displaying the editing of a standard tic-tac-toe rule.

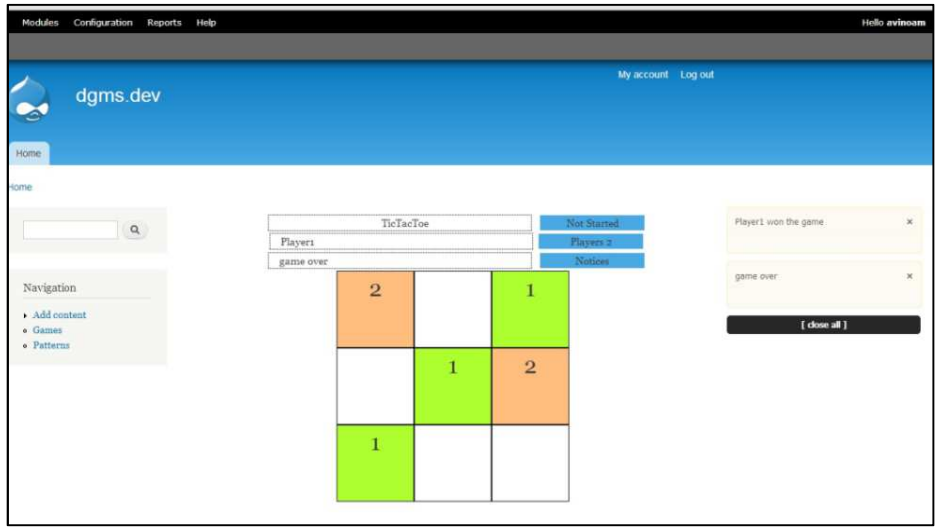

**Fig. 2.** Player Interface of the *GAMES* tool – This screen print displays the standard tic-tac-toe game, in which Player1 won the game. One can see the variable number of players – here the standard value of 2 Players – in the 2$^{nd}$ blue background small rectangle above the game board.

## 5   Game Case Studies

A series of games, serving as case studies was implemented to demonstrate our approach. Here we describe two such cases. First, Tic-Tac-Toe is discussed. Then, Sudoku a single player board game is described.

### 5.1   Tic-Tac-Toe and its Extensions

Besides the generic rules described in section 3, for each game there are specific rules. We first concisely discuss specific rules for the well-known tic-tac-toe standard game played on a three-by-three matrix game-board.

The specific rule that we which to point out is the check-winner rule. It can be stated in three different levels that may be checked to declare a winner:

- *Tile level* – in this level one specifies all the individual tiles that must be checked, this is the lowest and most tedious level;
- *Row*-Column-Diagonal level – in this intermediate level one specifies the rows, columns and diagonals that must be checked;
- *Overall Pattern* – in this highest level a single pattern composed of the previous level components is specified.

In Fig. 3 one sees the *GAMES* editor interface. One specifies in the Conditions area the Overall Pattern, composed of any row, any column and two matrix diagonals. It also displays two Actions: a- game set winner; b- game over.

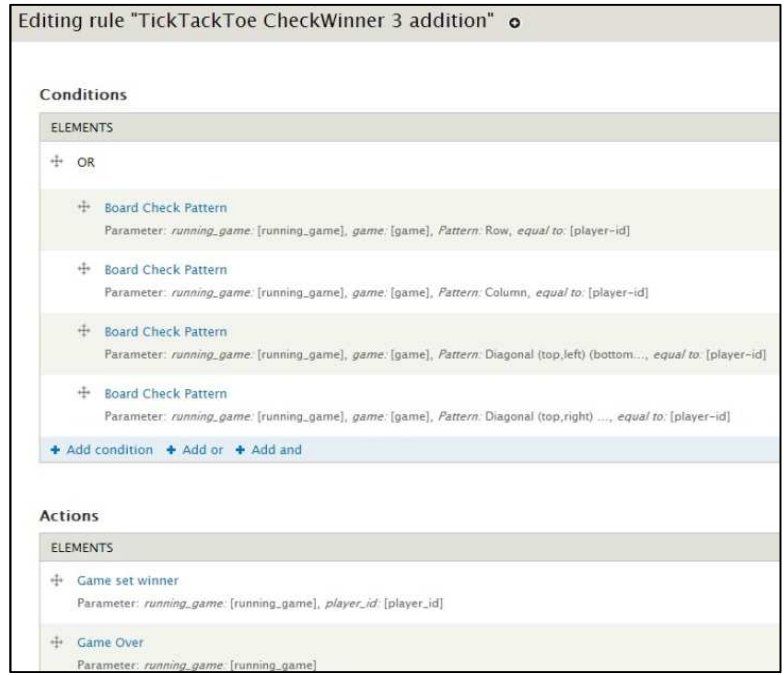

**Fig. 3.** Editor Interface of the *GAMES* tool – It displays the editing of the "Check Winner" rule for standard tic-tac-toe. It shows the Overall Pattern conditions (all rows, all columns, two diagonals) and two actions: Game set winner, and Game over.

Tic-Tac-Toe may be easily extended in several different ways, using the *GAMES* tool. The first conceivable extension is to change the size of the matrix. For instance, one can play with a four-by-four matrix. One could also extend the 2-dimensional to a 3-dimensional matrix.

Another extension is the change of the "Check Winner" rule. One could preserve the rule of a three-by-three matrix when using larger matrices. One could use a 4-tiles rule for four-by-four matrix. One could arbitrarily change the patterns – say instead of using just rows, columns and diagonals – one could use more complicated patterns in the plane or out of the plane in case of 3-dimensional matrices.

As a third conceivable extension, one could add more than 2 players, keeping their turns in round-robin fashion, or even by randomizing their turns.

### 5.2   Sudoku and its Extensions

Sudoku is interesting in our context, since it still is a board-game, but with a very different nature. Sudoku can be easily represented and is extensible within *GAMES* since it basically has common properties with other games such as tic-tac-toe:

- *Game Board* – it has 2-dimensional matrix as its basis;
- *Number of Players* – it needs just one player, but this can be extended;
- *Tile Values* – decimal digits, instead of the (0,1,2) tic-tac-toe values.

Fig. 4 shows the Sudoku player interface that is comparable to tic-tac-toe in Fig. 2.

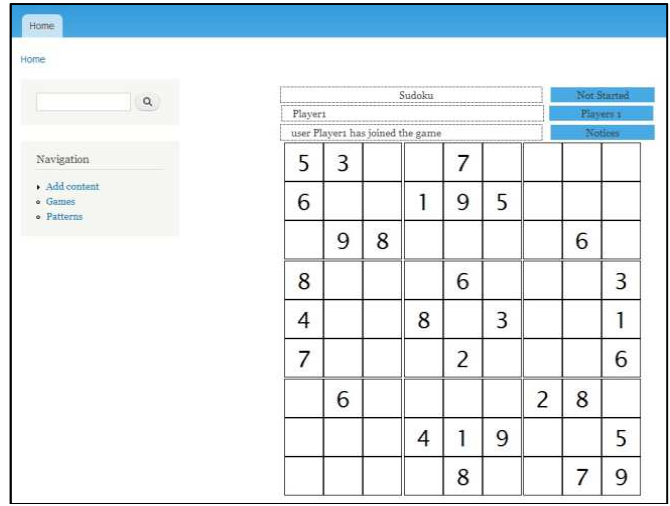

**Fig. 4.** Player Interface for Sudoku – This screen print displays the initial state – "Not Started" yet. The number of players is the standard one.

Before this work on extensible games, Sudoku has been extended in several ways by means of different sizes, shapes, and directions, use of inequalities and use of arithmetic operations, use of letters instead of digits. Some of the variants are Tatami Sudoku and Suguru (covering with different shapes and directions), Futoshiki (using inequalities), and Kakuro (sum operations). Besides these ones there are also western variations (e.g. Hidato). The *GAMES* tool also allows variable number of players.

## 6  Discussion

A knowledge-based infrastructure for game design and extension by non-programmers has been proposed. A tool was implemented upon the Drupal framework, using rules for games of increasing complexity, to demonstrate the approach.

### 6.1     Game Characterization and Extension Rationale

The game infrastructure has been shown to be generic enough to deal with quite different board games, such as tic-tac-toe, Sudoku and recently chess. Games have been characterized by sets of basic concepts, a state-machine and rules that make games playable. These are both of generic and specific types.

One[1] could ask: should anything forbid a user to modify tic-tac-toe into an adventure game? We believe not. The explanation lies in two possible motivations for game extension. The first motivation is entertainment. There is no reason to curb the tool power: we wish to stimulate free invention of totally new games. The second motivation is the investigation of limitations on game identification: when does a modified tic-tac-toe stop to be recognizable as such an extension? Again, we believe there is no reason to set bounds, acting as impediments to the investigation.

### 6.2   Software Engineering for Non-Programmers

The lessons learned from the domain of board games can be generalized to any application. The knowledge-driven approach for application design by non-programmers needs:

- *A domain ontology* – a set of generic concepts for the application domain;
- *A state-machine* – a set of states describing the potential behaviour of the application;
- *Transition rules* – governing the transitions among states; note that the rules' characteristics (event-condition-action) are equivalent to the properties of state transitions in Harel's statecharts.

Another observation is that sets of specific game moves can be generalized to specific scenario cases for general applications.

### 6.3   Future Work

Current work in progress includes extensions to more complex games such as chess.
Following the extension of tic-tac-toe to bigger boards and increasing the number of players, it is not difficult to perceive that above a certain number of players there is no possibility of winning the game. A very general – for any type of board game – and challenging issue is: how to determine an upper bound on number of players for given board size, which implies no winners. An even more general issue is: when does an extension cause a game to become unplayable?
Concerning game consistency and recognition, we propose the following: define a type of games – say sudokus – by a generic ontology. A Sudoku extension is said to belong to the type, if its specific ontology differs from the generic type ontology by some quantitative measure below a pre-determined threshold. A systematic investigation of suitable ontologies for game types is desirable.

### 6.4   Main Contribution

The main contribution of this work is the usage of knowledge-driven software tools for *game extension,* as a particular example of software engineering by non-programmers.

---

[1] We are grateful to an anonymous referee for raising these issues.

## References

1. Brom, C., Gemrot, J., Bida, M., Burkert, O., Partington, S.J. and Bryson, J.J.: POSH Tools for Game Agent Development by Students and Non-Programmers, in Proc. of CGAMES IEEE 2006, Dublin, Ireland (2006).
2. Exman, I.: A Non-concept is Not a ¬Concept, in Proc. KEOD International Conf. Knowledge Engineering and Ontology Development, pp. 401-404, (2012).
3. Good, J., and Robertson, J.: Computer Games Authored by Children: A Multi-Perspective Evaluation, in Proc. IDC 2004, pp. 123-124 (2004).
4. Hagen, U.: Where do Game Design Ideas Come From? Innovation and Recycling in Games Developed in Sweden, in Proc. DiGRA 2009 on Breaking New Ground: Innovation in Games, Play, Practice and Theory, (2009)
5. Johnson, W.L. and Beal, C.: Iterative evaluation of an intelligent game for language learning, in Proc. Of AIED 2005, Amsterdam, IOS Press, (2005).
6. McNaughton, M., Cutumisu, M., Szafron, D., Schaeffer, J., Redford, J. and Parker, D.: ScriptEase: Generative Design Patterns for Computer Role-Playing Games, in Proc. Automated Software Engineering, ASE'04, Linz, Austria, (2004).
7. Moreno-Ger, P., Sierra, J.L., Martinez-Ortiz, I. and Fernandez-Manjon, B.,: A Documental Approach to Adventure Game Development, Science of Computing Programming, Vol. 67, pp. 3-31 (2007).
8. Studer, R., Erisson, H., Gennari, J., Tu, S., Fensel, D. and Musen, M.: Ontologies and the Configuration of Problem-Solving Methods, in Proc 10$^{th}$ Banff Knowledge Acquisition for Knowledge-Based System Workshop (KAW'96), Banff, Canada, (1996).
9. Zagal, J.P. and Bruckman, A., 2008. The Game Ontology Project: Supporting Learning While Contributing Authentically to Game Studies, in Proc. ICLS'08 8th international conference on International conference for the learning sciences - Volume 2, pp. 499-506, (2008), see also: http://www.gameontology.com/index.php/Main_Page.
10. Zagal, J.P., Fernandez-Vara, C. and Mateas, M., 2008. Rounds, Levels and Waves – The Early Evolution of Game Play Segmentation, in Games and Culture, (2008).
11. Zagal, J.P., Mateas, M.,Fernandez-Vara, C., Hochhalter B. and Lichti, N., 2007. "Towards an Ontological Language for Game Analysis", in Proc. DiGRA 2005 on Changing Views – Worlds in Play, (2007).